\documentstyle[twocolumn,aps]{revtex}
\begin{document}

\title{Quantization ambiguity, ergodicity, and semiclassics}
\author{L. Kaplan\thanks{kaplan@physics.harvard.edu}
 \\Department of Physics and Society of Fellows,\\ Harvard
University, Cambridge, MA 02138
}
\maketitle

\begin{abstract}
A simple argument shows that eigenstates of a classically
ergodic system are individually ergodic on coarse-grained scales. This
has implications for the quantization ambiguity in
ergodic systems: the difference between alternative
quantizations is suppressed compared with the
$O(\hbar^2)$ ambiguity in the integrable case.
For two-dimensional ergodic systems in the high-energy
regime, individual
eigenstates are independent of the choice of quantization
procedure, in contrast with the regular case, where even the ordering
of eigenlevels is ambiguous. Surprisingly, semiclassical methods are shown
to be much more precise for chaotic than for integrable systems.
\end{abstract}

\vskip 0.1in

For many years, it has been widely recognized that ``quantizing"
a given classical system is inherently an ambiguous procedure, as a large
family of quantum Hamiltonians may have the same classical 
limit~\cite{dirac}. For
example, given the classical dynamics of a particle constrained to move on a
closed loop, different choices of boundary condition give rise to different
phases relating classical paths of different winding number. Knowledge
of these Aharonov-Bohm phases is of course necessary to construct a
semiclassical dynamics (which includes interference between classical
paths), and thus many semiclassical theories correspond to the same classical
dynamics. Physically, this $O(\hbar)$ or gauge ambiguity may be
associated with the possibility of varying the magnetic flux enclosed by the
loop.

This is not all, however: there are also many quantum theories,
differing at $O(\hbar^2)$ or higher in the Hamiltonian, which all have the 
same {\it semi}-classical limit. There are many ways of seeing this 
$O(\hbar^2)$ ambiguity; one of the simplest is to imagine 
making a canonical transformation on the classical phase space, applying
the canonical quantization prescription to the new coordinates, and
then transforming back to the original coordinate system. Generically,
one then obtains a new quantum Hamiltonian which differs from the original
by $O(\hbar^2)$ plus higher order terms:
\begin{equation}
\label{h1h0}
\hat H'=\hat H +\hbar^2 \hat A + \cdots \,,
\end{equation}
where the operator $\hat A$ has a well-defined classical
limit~\cite{wilkinson}. As classical dynamics is of course independent
of the choice of coordinate system, this implies that quantization
is inherently ambiguous at second order in $\hbar$. 

An even more striking case is that of a particle constrained
to move on a two-dimensional surface embedded in three-dimensional space.
Here, the non-trivial metric contained in the kinetic term gives rise
to obvious operator-ordering ambiguities in canonical quantization; this has
led to much discussion in the literature over whether a term proportional
to the local Gaussian curvature $R$ of the surface should be added
to the Hamiltonian, and if so, what the proportionality constant should be
(different prescriptions suggesting $\hbar^2R/8$, $\hbar^2R/6$,
and $\hbar^2R/12$ as the 
``correct" answer)~\cite{curv,ali,morecons}. In the path integral approach,
ambiguities at the same order arise in choosing how to incorporate
the metric into the kernel and in deciding at what point in
the infinitesimal time interval to evaluate functions of the
metric~\cite{pathint}. Physically, one may define
dynamics on a constraint surface through a limiting process, where the strength
of restoring forces causing the particle to live on the surface is taken
to infinity~\cite{maraner}. Classically, this procedure
is known to give an unambiguous constrained
dynamics\cite{arnold}; in quantum mechanics, $O(\hbar^2)$ differences
arise depending on the precise way in which the strength of the constraining
potential is taken to infinity at various places along the
surface~\cite{qconstr}. The
ambiguity here has a clear physical meaning: to determine the true
quantum mechanics one needs to know the mechanism through which the particle
is bound to the surface of constraint; it is not sufficient to know only
the intrinsic properties of the surface itself. Similarly, there is
no ``correct" answer to the problem of quantizing a classical
double pendulum~\cite{ali}:
quantum dynamics at $O(\hbar^2)$ is determined by the
precise way in which one takes to infinity the rigidity of the two rods.

In a two-dimensional system, the energy spacing between adjacent levels
is $O(\hbar^2)$, i.e. of the same order as the quantization ambiguity
demonstrated above. It would then seem not to be possible to uniquely
determine the eigenvalues and eigenstates of a two-dimensional system
given only the classical dynamics on the two-dimensional surface.
Similarly, it should not be generally possible to compute semiclassically
the levels and wavefunctions of a two-dimensional quantum system,
because a given semiclassical calculation has many quantum theories
corresponding to it, with Hamiltonians related as in Eq.~\ref{h1h0}.
Recently, however, it was shown (and numerically confirmed)
that in strongly chaotic systems long-time
semiclassical methods can in fact be used
to compute quantum properties to accuracy much better than a level spacing,
i.e. that individual wavefunctions and eigenenergies can be
semiclassically resolved~\cite{ltsc}.
This creates an apparent paradox, to be resolved
in the present paper.

The answer is somewhat surprising: it is found
that (i) the quantization ambiguity is greatly reduced for classically
ergodic as compared with regular systems; (ii) in two dimensions, the 
ambiguity for ergodic systems is small compared to a level spacing, in contrast
with the integrable case where even the ordering of eigenlevels is ambiguous;
and (iii) unexpectedly, semiclassical methods are valid to much longer times
in strongly chaotic as compared with integrable systems, allowing
individual eigenenergies to be easily resolved.
The key to these surprising results is the coarse-grained ergodicity
of individual wavefunctions in classically ergodic systems. More precisely,
in the classical limit, the quantum expectation value of any classically
defined operator over individual eigenstates must approach the ergodic,
microcanonical average of the operator, for almost all eigenstates. 
In effect, the fraction of eigenstates that deviate from ergodicity when
smoothed over a finite mesh size in phase space must tend to zero in the
$\hbar \to 0$ limit. This behavior has been studied mathematically under
a variety of technical assumptions by Shnirelman, Zelditch,
and Colin de Verdiere~\cite{szcdv}. Here we present a
simple physical argument that demonstrates the generality of the result.

Let $\hat A$ be an arbitrary quantum operator having the smooth
phase space function
$A(q,p)$ as its classical limit. Now we select any energy $E_0$
and choose a classically small energy window $\Delta E$ such that
the microcanonical average $a(E)$ of the function $A(q,p)$ is as close
as one likes to a constant, $a_0$, for all $E \in[E_0,E_0+\Delta E]$.
Notice that in the $\hbar \to 0$ limit, the classically infinitesimal
energy window will contain an arbitrarily large number
of quantum eigenstates, $N \sim \hbar^{-d}$ in $d$ dimensions.
We wish to show that for any $\epsilon$,
there exists
$\hbar$ such that
\begin{equation}
\label{sherg}
\left \langle
\left ( \langle n |\hat A|n \rangle - a_0 \right )^2 
\right \rangle< \epsilon^2\,,
\end{equation}
where the average is taken over all wavefunctions $|n \rangle$
in the energy window 
$[E_0,E_0+\Delta E]$.

Consider the quantum correlator 
\begin{equation}
f(t)={\rm tr} \; \hat A^\dagger \hat A(t) = {\rm tr} \;
\hat A^\dagger e^{i \hat H t} \hat A e^{-i \hat H t} \,,
\end{equation}
where the trace is over all states in the energy window.
Now define the time-averaged correlator:
\begin{eqnarray}
\label{ft}
F(T)&=&{1 \over \sqrt{2 \pi T}} \int dt \; e^{-t^2/2 T^2} f(t) \nonumber \\
&=& {1 \over N}
\sum_{n,n'=1}^N \left | \langle n |\hat A| n' \rangle \right |^2
e^{-(E_n-E_{n'})^2T^2/2} \,.
\end{eqnarray}
$F_{\rm cl}(T)$, the time-averaged correlator of $A(q,p)$ and
$A(q(t),p(t))$, is the classical counterpart of $F(T)$, and
must by the definition
of ergodicity tend to the ergodic value $a_0^2$ as $T \to \infty$.
Now we simply choose $T$ large enough so that $F_{\rm cl}(T)$ is as close
as we like to its long-time asymptotic value (say, within $O(\epsilon^2)$),
and then choose $\hbar$
small enough so that the Ehrenfest time $T_{\rm Ehr}$ (the time at which
classical--quantum correspondence breaks down) is large compared with
$T$. [Notice that for a hard chaotic system, the breakdown time
$T_{\rm Ehr} \sim \lambda^{-1}\log \hbar^{-1}$ as $\hbar \to 0$, where
$\lambda$ is the Lyapunov exponent; however we make no specific assumption
here about the system apart from its being ergodic.] 
We then have 
\begin{eqnarray}
\label{res}
{1 \over N}  \sum_{n=1}^N
\left | \langle n |\hat A|n \rangle \right |^2 \le F(\infty) \le F(T)
=a_0^2+O(\epsilon^2) \,.
\end{eqnarray}
Now $a_0$ is of course the mean value of the matrix elements
$\langle n |\hat A|n \rangle$ (which can be seen formally by considering
the correlation of $\hat A(t)$ with the identity operator), so Eq.~\ref{res}
implies that the variance of the matrix elements is bounded by $\epsilon^2$
(proving the statement of
Eq.~\ref{sherg}), and goes to zero in the $\hbar \to 0$ limit.

In the presence of symmetry-induced degeneracies, the argument above
clearly holds for {\it any} choice of orthonormal eigenbasis
$|n\rangle$. Furthermore,
the difference between the first two quantities in Eq.~\ref{res} is also
bounded above by $O(\epsilon^2)$; this implies
\begin{equation}
\langle n| \hat A |n' \rangle \to 0
\end{equation}
for almost all degenerate eigenstates  $|n\rangle$ and $|n'\rangle$
in the $\hbar \to 0$ limit.

As an aside, we mention that the constraint $F(\infty) \le F(T)$
(obtained from Eq.~\ref{ft}) allows us to circumvent the usual problem of
noncommutativity of the $\hbar \to 0$ and $T \to \infty$ limits.
Specifically, quantum approach to ergodicity at short times in the sense
of $F(T) \to a_0^2$ guarantees quantum ergodicity at longer times
where individual eigenstates are resolved. 

The statistical argument presented here leaves open the
possibility of a zero measure set of states grossly violating the
ergodic condition $\langle n | \hat A |n\rangle \approx a_0$. This is not a
deficiency in the argument: infinite sequences of highly non-ergodic states
(namely, the bouncing-ball wavefunctions) have been shown to exist
in chaotic systems with marginally stable classical orbits~\cite{bb}.
The fraction
of such states does go to zero in the classical high-energy limit, in
accordance with the predictions of the theorem.

We also note that no assumptions have been made in the argument
about ``hard chaos"
properties such as hyperbolicity or mixing; implications for the
elimination of quantization ambiguity in two dimensions thus hold for
all classically ergodic systems. 

Coarse-grained quantum ergodicity is a much weaker and more general
result than random matrix theory (RMT), which has been conjectured
to be a statistical description of classically ergodic
wavefunctions~\cite{rmtconj}. The latter conjecture suggests ergodic 
behavior of quantum wavefunctions on single-wavelength scales, and is violated
by scars~\cite{scars} and other effects of short-time
dynamics~\cite{wqe,sinai}.
For example, the quantum wavefunctions of the Sinai billiard, a paradigm
of classical chaos, have been shown to have inverse participation ratios
(IPR's) that diverge in the classical limit away from their ergodic and RMT
values~\cite{sinai}. This strongly non-ergodic  wavefunction behavior 
on scales of size $\hbar$ is entirely consitent with ergodicity on
coarse-grained scales as discussed above. It is easy to see that
the ergodicity argument breaks down if $\hat A$ does not have a well-defined
classical limit. We may for example let $\hat A$ be a projection onto a
localized test state $\hat A = |a \rangle \langle a|$, but then we are
not guaranteed to find a time $T$ where classical--quantum correspondence
between $F(T)$ and $F_{\rm cl}(T)$ still holds, and simultaneously
$T \ll T_{\rm Ehr}$.

Comparing the result of Eq.~\ref{sherg} with Eq.~\ref{h1h0}, we see that the 
leading effect of a change in the quantization prescription is to shift
all energy levels classically close to 
$E_0$ by a constant displacement $\hbar^2a_0$
(comparable in size to a level spacing);
perturbation theory then gives:
\begin{equation}
\label{delen}
\delta E_n \equiv E'_n - E_n = \hbar^2(a_0+O(\epsilon))
\end{equation}
with $\epsilon \ll 1$ for small $\hbar$ (high energy). In the special case
of RMT, the energy hypersurface has $M \sim \hbar^{1-d}$ Planck-sized cells:
since the deviations $\epsilon$ in the matrix elements
$\langle n|\hat A|n\rangle$ arise from uncorrelated fluctuations in these
cells, we then have
\begin{equation}
\epsilon_{\rm RMT} \sim {1 \over \sqrt M} \sim \hbar^{(d-1)/2} \,.
\end{equation}
The relative shift between nearby energy levels compared to a level
spacing is
\begin{equation}
\label{enshift}
{\delta E_n - \delta E_m \over \hbar^d} \sim
\hbar^{2-d} \epsilon  \;
{\buildrel {\rm \tiny RMT} \over \sim}  \; \hbar^{(3-d)/2} \,.
\end{equation}

The overall energy shift $\hbar^2 a_0$ is of course unphysical,
corresponding simply to changing the Hamiltonian by a constant;
only energy differences are measurable, and from
the first relation in Eq.~\ref{enshift} we see that for $d=2$
these become independent of one's choice of quantization in
the $\hbar \to 0$ limit (i.e. for highly excited eigenstates).
Energy splittings $E_n-E_m$ which are already classically large can
of course change by $O(\hbar^2)$ as one considers different quantum
systems with the same semiclassical limit. However, the ratio
$\delta(E_n-E_m)/(E_n-E_m) \to 0$ for almost all states $|n\rangle$,
$|m\rangle$ in the semiclassical limit for $d=2$, whether 
$|n\rangle$ and $|m\rangle$ are nearest neighbor levels or are far apart
in energy. Thus, in the classical limit of a $d=2$ constrained surface,
the way in which the particle is bound to the surface has no effect
on any measurable quantity, {\it provided that motion on the constraint
surface itself is ergodic}. 
[In the case of mixed phase space, states living on regular islands
move up or down relative to the rigid chaotic sea, as the quantization
is varied.]

The result of Eq.~\ref{enshift} is consistent
with the finding in \cite{ltsc} that in caustic-free chaotic $d=2$ systems,
long-time semiclassical dynamics approaches the quantum answer, and
that the critical dimension for breakdown of the semiclassical
approximation at the Heisenberg time is $d=3$ for chaotic systems,
as compared with $d=2$ in the integrable case.

For a simple numerical example of the above results, we consider a
discrete-time map on a two-dimensional toroidal phase space 
$(q,p)\in [0,1) \times [0,1)$ (notice that
this scales equivalently to a two-dimensional autonomous Hamiltonian system
and can be thought of as a Poincar\'e section of the latter):
\begin{eqnarray}
\label{eqmo}
p \to \tilde p &=& p - V'(q) \; {\rm mod} \; 1 \nonumber \\
q \to \tilde q &=& q + \tilde p \; {\rm mod} \; 1 \,.
\end{eqnarray}
The above equations of motion can be obtained from the
time-periodic (``kicked") Hamiltonian
\begin{equation}
H={p^2 \over 2} + V(q)\sum_n \delta(t-n)\,.
\end{equation}
We let the potential be given by
\begin{equation}
V(q)=\pm { q^2 \over 2} + {0.4 \over (2\pi)^2} \sin{2 \pi q}
+r h^2 \cos{2 \pi q}\,,
\end{equation}
where the $-$ sign gives completely chaotic dynamics~\cite{ltsc},
while the $+$ sign leads to a (mostly) regular classical phase space.
The $\sin{2 \pi q}$ term is added to break parity symmetries and make
the dynamics nonlinear and generic in both cases, while $r$
parametrizes a one-parameter family of possible quantizations.

A sample piece of the regular (solid curve) and chaotic (dashed
curve) spectrum is shown in Fig.~1, as a function of the 
quantization parameter $r$. We see in the regular case that the
eigenvalues often cross each other as the quantization is varied,
making impossible a semiclassical ordering of the spectrum, while
in the chaotic case the eigenvalues never shift by an amount
comparable to a mean level spacing ($\approx 0.0117$ in this
calculation). We notice that in our example the trace of $A(q,p)=
r \cos{2 \pi q}$ is zero, so the overall spectral shift 
$\hbar^2 a_0$ of Eq.~\ref{delen} vanishes as well. For a different
$A(q,p)=r \left[\cos{2 \pi q}\pm c \right]$,
there would be a secular upward or downward trend
in all the eigenvalues (regular and chaotic) as $r$ increased, but
no physical quantities would be affected.

Despite the exponential
proliferation of classical paths in the chaotic case,
semiclassical calculations past the Heisenberg time can be performed
to arbitrary precision with only a power-law amount of effort,
using an iterative approach,
and individual semiclassical eigenvalues and eigenstates can be
extracted~\cite{ltsc}. Semiclassically obtained eigenvalues for the 
chaotic system are indicated by arrows in Fig.~1 and are observed to
agree well with the quantum results (independent of quantization). In
the regular case, there is of course little meaning to semiclassically
computed individual eigenvalues, and these are not shown.

In Fig.~2 we show the mean squared change in eigenlevel position,
as a function of effective Planck's constant (inverse momentum)
$h=1/64 \cdots 1/1028$,
when the quantization parameter $r$ is changed from 
$0.0$ to $3.0$. The $\times$'s represent the regular case,
and show that the quantization ambiguity there is large and energy-independent.
The data points marked by squares correspond to the chaotic system,
and clearly follow the linear law predicted by Eq.~\ref{enshift}
(for $d=2$). Finally, the plusses come from the ``slow ergodic"
sawtooth potential map~\cite{wqe} defined by
\begin{equation}
V(q)=0.3\left|q-{1 \over 2} \right|\,,
\end{equation}
where the quantization ambiguity is now put in the kinetic term:
$H=\cdots + rh^2\cos{2 \pi p}$. Here we do not expect quantum wavefunctions to
be ergodic on the scale of a single channel in momentum space~\cite{wqe},
but they will be ergodic on scales $\sim 1/|\log \hbar|$ as $\hbar \to 0$.
We then expect $\epsilon \sim 1/(\alpha+\beta \log \hbar)$ in
Eq.~\ref{enshift}, which behavior is indeed observed in Fig.~2. We see that
ergodicity without chaos is sufficient to obtain unambiguous quantization
of $d=2$ systems, though the error in the high-energy limit
approaches zero more slowly in the slow ergodic than in the fully
hyperbolic case.

It is a pleasure to thank M. Wilkinson for initially stimulating this
work
and E. J. Heller for many fruitful discussions.
This research was supported by the National Science Foundation under
Grant No. 66-701-7557-2-30.

\begin{figure}
\caption{Sample spectrum as a quantization parameter is varied: solid line
represents the regular and dashed line the chaotic case, both for 
$h=1/512$. Arrows indicate
eigenvalues of the long-time semiclassical chaotic dynamics.}
\label{fig1}
\end{figure}

\begin{figure}
\caption{Mean squared change in eigenvalues between two quantizations
(in units where mean
level spacing $=1$), plotted as a function of the effective Planck's constant.}
\label{fig2}
\end{figure}


\begin{references}
 
\bibitem{dirac} P.\,A.\,M.\,Dirac, {\it Lectures on Quantum Mechanics},
Academic Press, New York (1967).

\bibitem{wilkinson} M. Wilkinson, {\it J. Phys.} {\bf A 21}, 1173 (1988).

\bibitem{curv} M.\,C.\,Gutzwiller, {\it  Chaos in Classical and
Quantum
Mechanics}, Springer-Verlag, New York (1990).

\bibitem{ali} M.\,K.\,Ali, {\it Can.\,J.\,Phys.} {\bf 74}, 255 (1996).

\bibitem{morecons} A few key papers on the vast subject of
constrained quantization include:
K. S. Cheng, {\it J. Math. Phys.} {\bf 13}, 1723 (1972);
B. S. Dewitt, {\it Phys. Rev.} {\bf 85}, 653 (1952);
{\it Rev. Mod. Phys.}
{\bf 29}, 3774
(1957);
I.\,A.\,Batalin, E.\,S.\,Fradkin, and T.\,E.\,Fradkina, {\it Nucl. Phys.}
{\bf B 332}, 723 (1990);
I.\,A.\,Batalin and E.\,S.\,Fradkin, {\it Nucl. Phys.} {\bf B 326},
701 (1989);
L.\,D.\,Faddeev, {\it Theor. Math. Phys.} {\bf 1}, 1 (1970);
J.\,Sniatycki, {\it Geometric Quantization and Quantum Mechanics,} Springer,
New York (1980);
T.\,Kimura, {\it Prog. Theor. Phys.} {\bf 46}, 1261 (1971);
R.\,C.\,T.\, da Costa, {\it Phys. Rev.} {\bf A 23}, 1982 (1981);
N.\,T.\,De Oliveira and R.\,Lobo, {\it Nuovo Cimento} {\bf 71B},
196 (1982).

\bibitem{pathint} L.\,S.\,Schulman, {\it Techniques and Applications of
Path
Integration}, John Wiley and Sons, New York (1981).

\bibitem{maraner} P. Maraner, {\it J. Phys.} {\bf A 28}, 2939 (1995);
P. Maraner and C. Destri, {\it Mod. Phys. Lett.} {\bf A 8}, 861 (1993).

\bibitem{arnold} V.\,I.\,Arnol'd,
{\it Mathematical Methods of Classical Mechanics}, Springer-Verlag, New
York (1978); C. Lanczos, {\it The Variational Principles of Mechanics},
Dover (1986).

\bibitem{qconstr} L. Kaplan, N. T. Maitra, and E. J. Heller,
{\it Phys. Rev.} {\bf A 56}, 2592 (1997).

\bibitem{ltsc} L. Kaplan, {\it Phys. Rev.} {\bf E 58}, 2983 (1998);
L. Kaplan, {\it Phys. Rev. Lett.} {\bf 81}, 3371 (1998).

\bibitem{szcdv} A. I. Shnirelman, {\it Usp. Mat. Nauk.} {\bf 29}, 181
(1974); Y. Colin de Verdiere, {\it Commun. Math. Phys.}  {\bf 102}, 497 (1985);
S. Zelditch, {\it Duke Math. J.} {\bf 55}, 919 (1987);
S. Zelditch and M. Zworski, {\it Commun. Math. Phys.} {\bf 175}, 673 (1996).

\bibitem{bb} P. W. O'Connor and E. J. Heller,
{\it Phys. Rev. Lett.} {\bf 61}, 2288 (1989).

\bibitem{rmtconj} M. V. Berry, in {\it Chaotic Behaviour of
Deterministic Systems}, ed. by G. Iooss, R. Helleman, and R. Stora
(North-Holland
1983) p. 171; O. Bohigas, M.-J. Giannoni, and C. Schmit, {\it J. Physique
Lett.} {\bf 45}, L-1015 (1984).

\bibitem{scars} E. J. Heller,  {\it Phys. Rev. Lett. } {\bf 53}, 1515 (1984);
L. Kaplan, {\it Phys. Rev. Lett.} {\bf 80}, 2582 (1998);
L. Kaplan, {\it Nonlinearity} {\bf 12}, R1 (1999).

\bibitem{wqe} L. Kaplan and E. J. Heller,
{\it Physica} {\bf D 121}, 1 (1998).

\bibitem{sinai} L. Kaplan and E. J. Heller, {\it 
Short-Time Effects on Eigenstate Structure in Sinai Billiards and
Related Systems}, in preparation (1999).

\end{references}
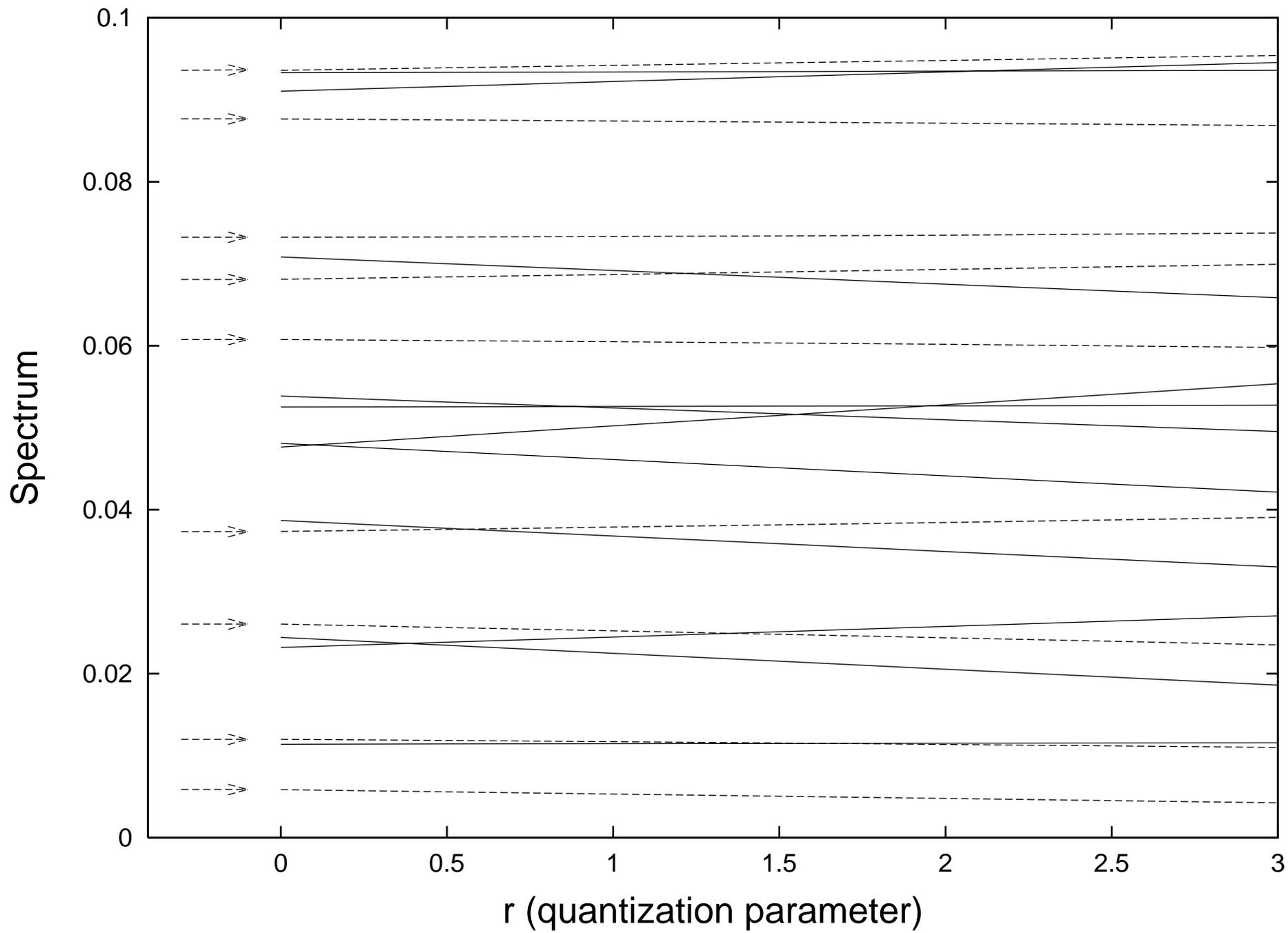
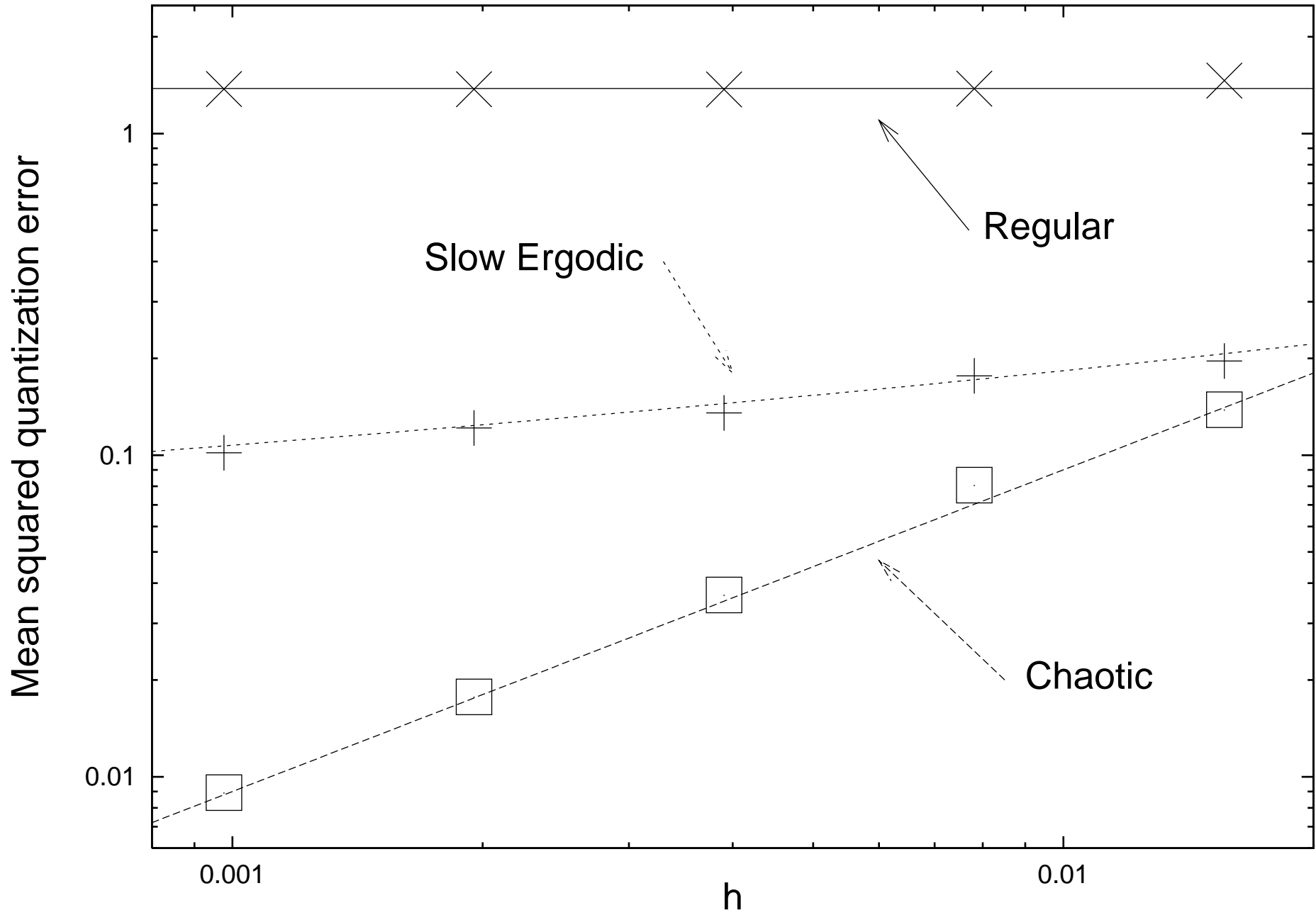
\end{document}